\documentclass[prl,aps,nofootinbib,showkeys,showpacs,twocolumn]{revtex4-1}   
\usepackage{epsfig}  
\usepackage{graphicx}  
\usepackage{amssymb}
\usepackage{amsmath}
\usepackage{color}  
\definecolor{darkgreen}{rgb}{0.2,0.5, 0.2}

\usepackage{dsfont}

\begin{document}

\title{  Effect of shell structure on the fission of sub-lead nuclei}

\author{Guillaume Scamps}
\email{gscamps@ulb.ac.be}
\affiliation{Center for Computational Sciences, 
University of Tsukuba, Tsukuba  305-8571, Japan}
\affiliation{Institut d’Astronomie et d’Astrophysique, Universit\' Libre de Bruxelles, Campus de la Plaine CP 226, BE-1050 Brussels, Belgium}
 \author{C\'edric Simenel}  
 \email{cedric.simenel@anu.edu.au}  
\affiliation{Department of Theoretical Physics and Department of Nuclear Physics, Research School of Physics and Engineering \\ Australian National University, Canberra, Australian Capital Territory 2601, Australia}

%


%
%

\begin{abstract}
Fission of atomic nuclei often produce mass asymmetric fragments. 
However, the origin of this asymmetry was believed to be different in actinides and in the sub-lead region [A. Andreyev {\it et al.}, Phys. Rev. Lett. {\bf 105}, 252502 (2010)]. 
It has recently been argued that quantum shell effects stabilising pear shapes of the fission fragments could explain the observed asymmetries in fission of actinides
 [G. Scamps and C. Simenel, Nature {\bf 564}, 382 (2018)]. 
This interpretation is tested in the sub-lead region using 
microscopic mean-field calculations of fission based on the Hartree-Fock approach with BCS pairing correlations.
The evolution of the number of protons and neutrons in asymmetric fragments of mercury isotope fissions is  interpreted in terms of deformed shell gaps in the fragments. A new method is proposed to investigate the dominant shell effects in the pre-fragments at scission.
We conclude that the mechanisms responsible for asymmetric fissions in the sub-lead region are the same as in the actinide region, which is a strong indication of their universality. 
\end{abstract}

\maketitle


Nuclear fission was discovered in 1938  by bombarding $^{235}$U with slow neutrons, 
producing a heavy fragment in the barium region and a light one near krypton \cite{meitner1939,hahn1939}.
Most of the questions opened by this fundamental discovery are still debated 80 years later \cite{schmidt2018,andreyev2018}. 
In particular, the origin of the observed mass asymmetry between the fragments is  interesting 
as it offers a unique signature of quantum effects in large amplitude collective motion. 

The spherical shell model developed by Goeppert Mayer in 1950 \cite{mayer1950} 
 explains the extra stability of nuclei with so-called ``magic'' numbers of protons and neutrons 
associated with fully occupied quantum shells (analogous to noble gas in atomic physics).
Closed shells have then been naturally  invoked as possible drivers to asymmetric fission \cite{mayer1948,meitner1950,faissner1964,zhang2016,sadhukhan2017}, 
energetically favouring the formation of fragments with (doubly)magic clusters such as $^{132}_{\,\,50}$Sn$_{82}$. 
Neutron deformed shell effects in fission fragments with $N\approx88$ neutrons \cite{wilkins1976} as well as in the fissioning nucleus \cite{gustafsson1971} have also been invoked.
However, experiments show that the main driver to asymmetric fission in the actinide region 
is the number of protons of the heavy fragment, which remains particularly stable around $Z\approx54$ \cite{unik1974,schmidt2000,bockstiegel2008}.

We recently proposed a possible explanation for this stability~\cite{scamps2018} 
based on octupole (pear shape) deformed shell effects in the $^{144}$Ba region \cite{bucher2016,bucher2017}. 
The latter are induced by energy gaps at $Z=52$ and $56$ for a combination of quadrupole (cigar shape) and octupole deformations \cite{leander1985}.
Just before scission, the prefragments are connected by a neck which enforces their octupole deformation 
due to a combination of short-range nuclear attraction and long-range Coulomb repulsion. 
The production of nuclei like $^{144}$Ba which can exhibit octupole shapes for no or little cost in energy is then naturally favoured 
(unlike $^{132}$Sn which is hard to deform). 
This mechanism offers an explanation for mass asymmetric fission in actinides. 
However,  the question of its universality remains open, i.e., can it explain asymmetric fission in other regions of the nuclear chart? 

A new region of asymmetric fission has been discovered more recently in the sub-lead region \cite{andreyev2010} 
and actively studied experimentally by several groups \cite{ghys2014,nishio2015,prasad2015,tripathi2015,martin2015,rodriguez2016}.
In particular, $^{180}$Hg was found to fission asymmetrically, with heavy and light fragment mass distributions centred around $A\approx100$ and 80 nucleons, 
respectively, while its fission was expected to be symmetric due to closed spherical shells in $^{90}_{40}$Zr$_{50}$~\cite{andreyev2010}. 
It was then referred to as a ``new type of asymmetric fission'', as the observed asymmetry could clearly not be explained by spherical shell effects. 
Theoretically, this could reflect the presence of an asymmetric saddle-point with a ridge between symmetric and asymmetric fission valleys \cite{andreyev2010,ichikawa2012,moller2012,schmitt2017}.
Different explanations were proposed, involving shell effects in pre-scission configurations associated with dinuclear structures \cite{warda2012b}, or
with quadrupole deformed neutron shells in the fragments \cite{panebianco2012,andreev2012,andreev2013,andreev2016,Lemaitre2019} as well as in the fissioning nucleus \cite{ichikawa2019}.

In this letter, we show that the mechanism based on octupole deformed shell effects in the fragments, 
that we invoked to interpret asymmetric fission in actinides \cite{scamps2018}, also plays an important role in the sub-lead region. 
We first focus our theoretical analysis on $^{180}$Hg. 
We then investigate the evolution of asymmetric fission across the mercury isotopic chain. 
We finally present a comparison of predicted asymmetric modes with existing experimental data in the sub-lead region.


Our theoretical analysis is based on the Hartree-Fock (HF) self-consistent mean-field theory (or energy density functional approach) 
with BCS pairing correlations. 
This microscopic approach has been successfully used by several groups with various levels of sophistication to investigate 
both static \cite{flocard1974,warda2002,bonneau2006,dubray2008,pei2009a,staszczak2009,mcdonnell2013,mcdonnell2014,bernard2019} 
and dynamical \cite{simenel2014a,scamps2015a,tanimura2015,goddard2015,goddard2016,bulgac2016,simenel2018,bulgac2018,scamps2018} 
characteristics of fissioning nuclei (see also Ref.~\cite{schunck2016} for a recent review).

\begin{figure}[htb]
\begin{center}
\includegraphics[width= 1 \linewidth]{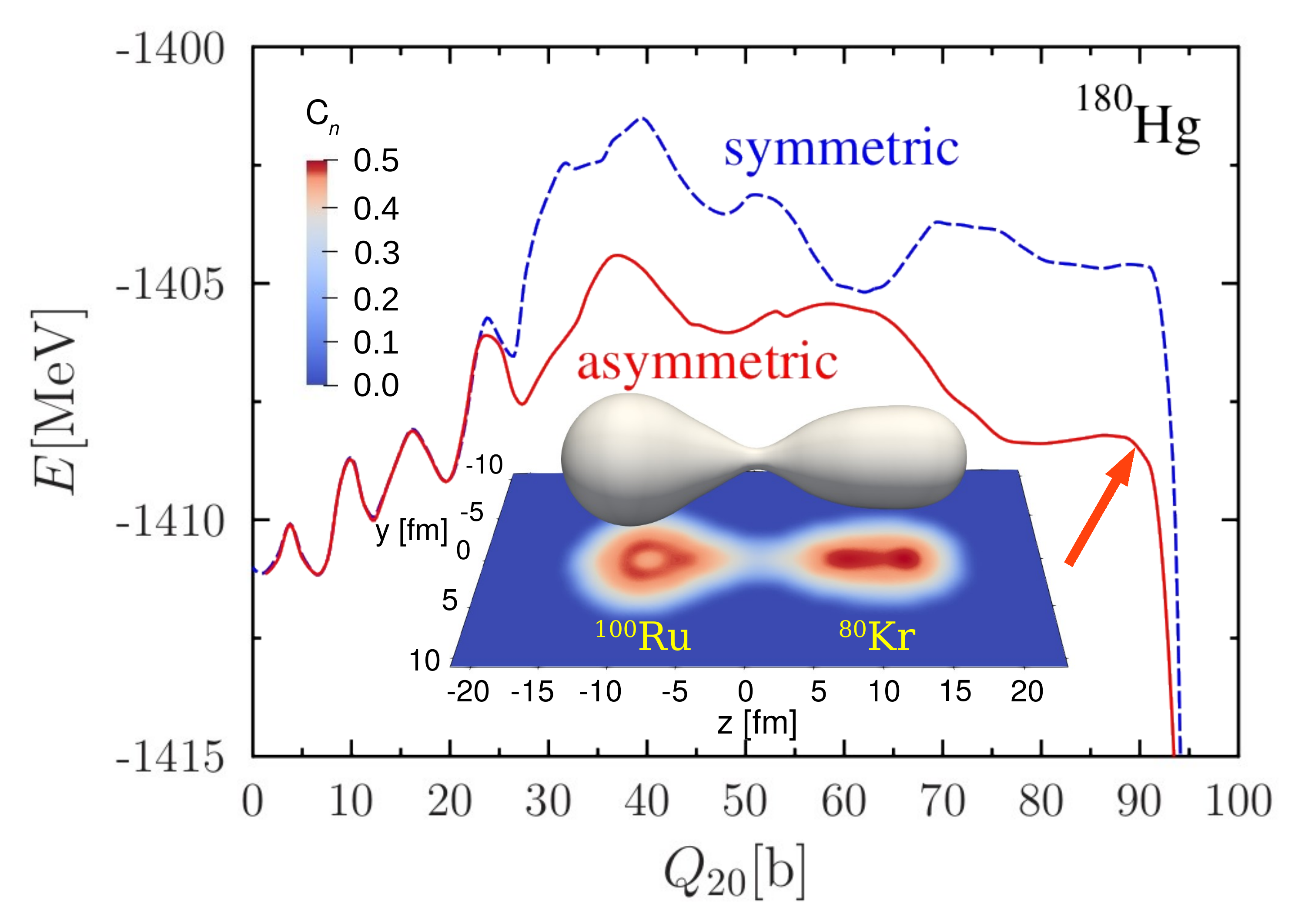}
\end{center}
\caption{ Potential energy as a function of quadrupole moment along the symmetric fission path (dashed line) and asymmetric fission valley (solid line)  of $^{180}$Hg. Iso-densitiy surface at half saturation density $\rho_0/2=0.08$~fm$^{-3}$ and neutron localisation function (projection) are shown for  
 $^{180}$Hg with quadrupole moment $Q_{20}=90$~b, just before scission in the asymmetric fission valley, as indicated by the red arrow.} 
\label{fig:180Hg}
\end{figure}

Here we choose a static approach.
 Indeed, for the sub-lead nuclei studied here, we do not expect dynamical effects to play a major role
as their fission valleys do not exhibit  long descent of the potential from saddle point to scission (see Fig.~\ref{fig:180Hg} and supplemental material Fig.~2). 
This is at variance with the actinide region in which 
dynamical effects occurring during the descent of the potential from saddle to scission 
may impact the outcome of the fission process (e.g., the total kinetic energy of the fragments \cite{simenel2014a}). 
Nevertheless, predictions of time-dependent and static approaches are in relatively good agreement 
in terms of the fragment mass and charge asymmetry.
Furthermore, for the systems we have studied in the sub-lead region, we found that the saddle and scission points are much closer than in actinides.
Thus, the dynamical effects are not expected to induce strong deviations from results obtained in a quasistatic picture.
We therefore conclude that the mechanisms responsible for mass-asymmetric fission in the sub-lead region can be studied with a time-independent microscopic approach. 

The constrained Hartree-Fock method with the BCS approximation for the pairing correlations is used. 
The nuclear interaction is described by the SLy4d parametrisation of the Skyrme energy density functional (EDF) \cite{kim1997}, 
with a surface type pairing functional with interaction strength $V_0^{nn}$ = 1256~MeV$\cdot$fm$^3$  and $V^{pp}$ = 1462~MeV$\cdot$fm$^3$ \cite{scamps2013a}. 
The calculations are done with a modified version of the  \textsc{ev8} solver \cite{bonche2005} where only one plane of symmetry is used. 
A spatial grid of dimension $L_x \times L_y \times 2L_z = 40\times 19.2 \times 19.2$~fm$^3$ with a mesh spacing of 0.8~fm is used. 

In order to study a large number of systems with  a unique method, the following procedure has been applied. 
First, a constrained calculation is done with a quadrupole constraint $Q_{20}=47.3$~b and several octupole constraints from $Q_{30}=0$ to 22~b$^{3/2}$. 
The octupole constraint is then released letting the system explore the bottom of the  fission valley(s). 
Then the valley potential energy curves are determined by making small evolutions of the quadrupole constraint 
from $Q_{20}$= 47.3 b to 0 and from $Q_{20}$= 47.3 b to 110 b.

It is also important to note that our fission valleys are obtained at zero temperature.
Modification of the potential energy surface at finite temperature can indeed have an effect on the asymmetry of the fragment mass distribution \cite{tao2017,zhao2019a,zhao2019b}. 
However, self-consistent calculations indicate that fission modes are expected to be weakly influenced by excitation energy in the mercury region \cite{mcdonnell2014}.

This approach allows us to study the shape of a system undergoing fission, and in particular to investigate the role of shell effects in the fragments. 
We also use it to predict the average number of protons and neutrons in the fragments for a given fission mode. 
A direct comparison can then be made with the centroids of experimental fission fragment mass distributions for asymmetric modes. 
Widths and shapes of these distributions, as well as a quantitative study of the competition between symmetric and asymmetric modes are beyond the scope of this work 
and would require a more advanced treatment of fluctuations via, e.g., 
the time-dependent generator coordinate method \cite{goutte2005,tao2017,zhao2019a,zhao2019b,regnier2018b,regnier2019}, 
stochastic dynamics on top of a potential energy surface \cite{moller2012,moller2015,sadhukhan2016}, 
or stochastic mean-field calculations \cite{tanimura2017}.


Figure~\ref{fig:180Hg} shows the evolution of the potential energy as a function of quadrupole moment for a symmetric path and in the asymmetric valley. 
The fact that the asymmetric valley remains significantly lower in energy than the symmetric path 
is a clear indication that asymmetric fission is energetically favoured in this system. 
The predicted outcome of the $^{180}$Hg asymmetric fission is a light fragment centred around 
$^{80}_{36}$Kr$_{44}$ and a heavy one around $^{100}_{\,\,\,44}$Ru$_{56}$, 
in excellent agreement with the masses observed experimentally~\cite{andreyev2010}.

The isodensity surface plotted in Fig.~\ref{fig:180Hg} shows that both fragments have  significant quadrupole and octupole deformations. 
The neutron localisation function \cite{becke1990,reinhard2011} (see supplemental material) which is shown as a projection in Fig.~\ref{fig:180Hg} 
also exhibits strong quadrupole and octupole shapes within the prefragments.
At scission, the associated quadrupole deformation parameters are $\beta_2^L\simeq 0.75$ 
and  $\beta_2^H\simeq 0.25$ for  $^{80}$Kr  and  $^{100}$Ru fragments, respectively, 
indicating a  compact heavy fragment and an elongated light one.
 The octupole deformation parameters are $\beta_3^L\simeq\beta_3^H\simeq 0.25$ for both fragments. 
These octupole  deformations are not surprising, as pear shapes are induced by the neck in which strong nuclear attraction between the prefragments is still present. 
Therefore, the formation of fragments with small octupole deformation energy (and even stable octupole shapes) is expected to be energetically favoured.
Conversely, the formation of nuclei which are hard to deform should be hindered.

\begin{figure}[htb]
\begin{center}
\includegraphics[width= 8cm]{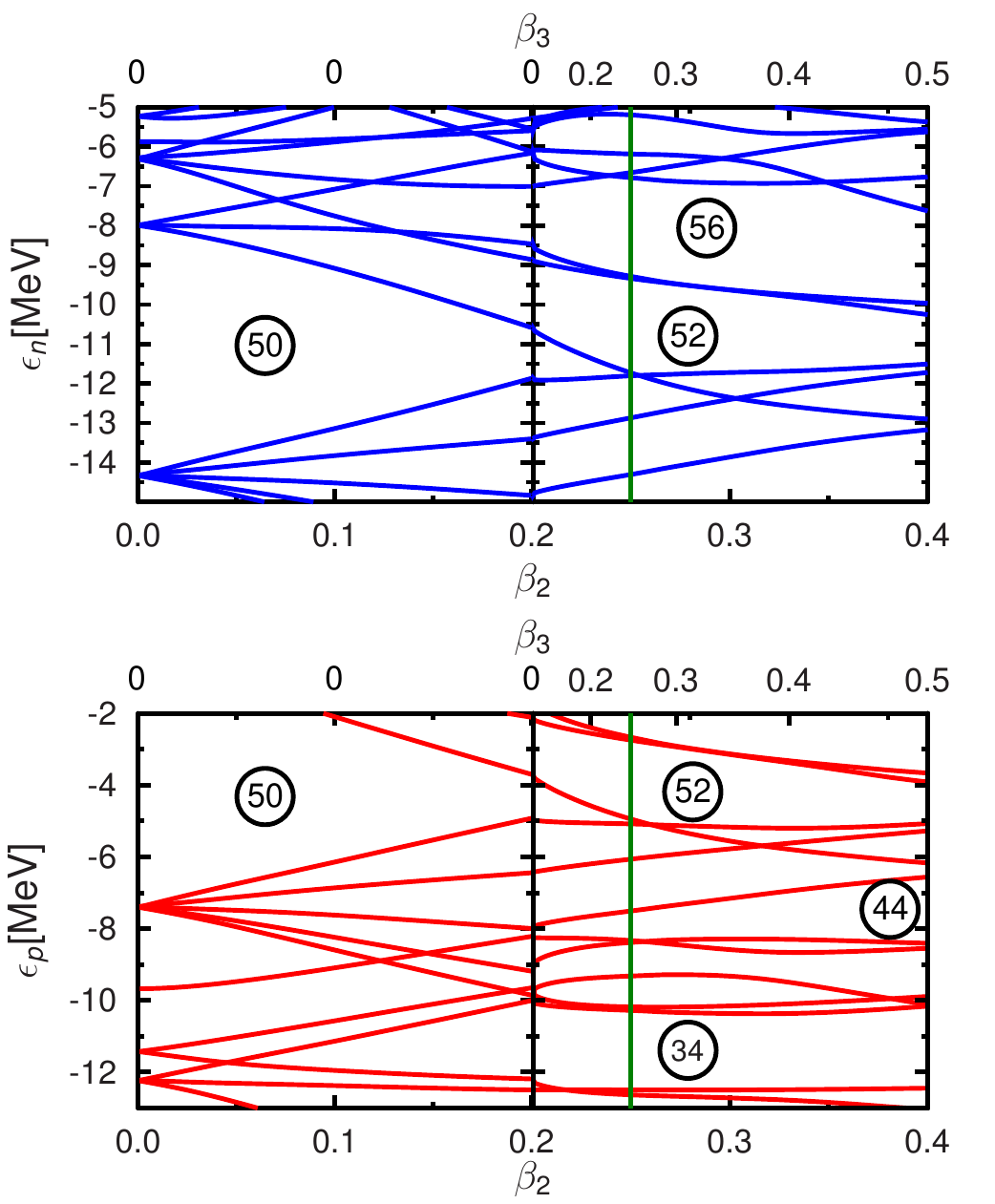}
\end{center}
\caption{ Neutron (top) and proton (bottom) single-particle energies as a function of the quadrupole (lower scale)  and octupole (upper scale)  deformation parameters in $^{100}$Ru (heavy fragment in $^{180}$Hg fission).
The left panel is obtained with an octupole constraint $\beta_3=0$.
The right panel is obtained by constraining $\beta_3$ and without constraint on $\beta_2$. 
} 
\label{fig:struc_heavy}
\end{figure}

HF+BCS calculations have been performed with quadrupole and octupole deformation constraints to investigate deformed shell effects in the fragments. 
Figure~\ref{fig:struc_heavy} shows the evolution of single-particle energies $\epsilon_{n}$ and $\epsilon_{p}$
 in $^{100}$Ru.  
The left panel is obtained by varying $\beta_2$ and with $\beta_3=0$.
The right panel is obtained by varying $\beta_3$ without  quadrupole constraint. 
In this case, energy minimisation with increasing $\beta_3$ leads to increasing $\beta_2$ as well (Fig.~\ref{fig:struc_heavy}-right).
Spherical gaps (``magic numbers'') at $\beta_{2,3}=0$  disappear with quadrupole deformation, 
while new deformed gaps appear.
In particular, large gaps at proton and neutron numbers 52 and 56 (compact fragment) are observed at the quadrupole and octupole deformations of the fragments at scission (green solid lines).

These shell gaps are present for a broad range of octupole deformations thanks to octupole correlations. 
The origin of octupole correlations in nuclei has been widely discussed in the litterature~\cite{butler1996,robledo2011,butler2016}.
In fact, these octupole correlations for $Z=52$ and $56$ protons \cite{leander1985} are expected to favour the formation of the heavy fragment (and thus drive asymmetry) in fission of actinides \cite{scamps2018}.
Comparing these numbers with those of neutrons in the fission fragments of $^{180}$Hg, 
we see that the large gap $N=56$ (Fig.~\ref{fig:struc_heavy}) is indeed
expected to favour the formation of $^{100}_{\,\,\,44}$Ru$_{56}$,
and can therefore explain the asymmetry observed in $^{180}$Hg fission. 

Deformed shell effects could also be present in the light fragment. 
We see in Fig.~\ref{fig:180Hg} that the $^{80}$Kr pre-fragment is strongly elongated ($\beta_2^L\simeq0.75$), in addition to its octupole deformation.
Shell gaps are often observed in single-particle spectra at large $\beta_2$, e.g., for $32-36$ or $42-46$ protons or neutrons (see, eg, \cite{nazarewicz1985} and supplemental material). 
However, to affect fission, these gaps must also be stable against a range of octupole deformations. 
The method used above to identify shell gaps using an isolated deformed fragment can be difficult to apply for large deformations as a combination of octupole and large quadrupole constraints reproduce the shape of the pre-fragment only approximatively.

\begin{figure}[htb]
\begin{center}
\includegraphics[width= 7cm]{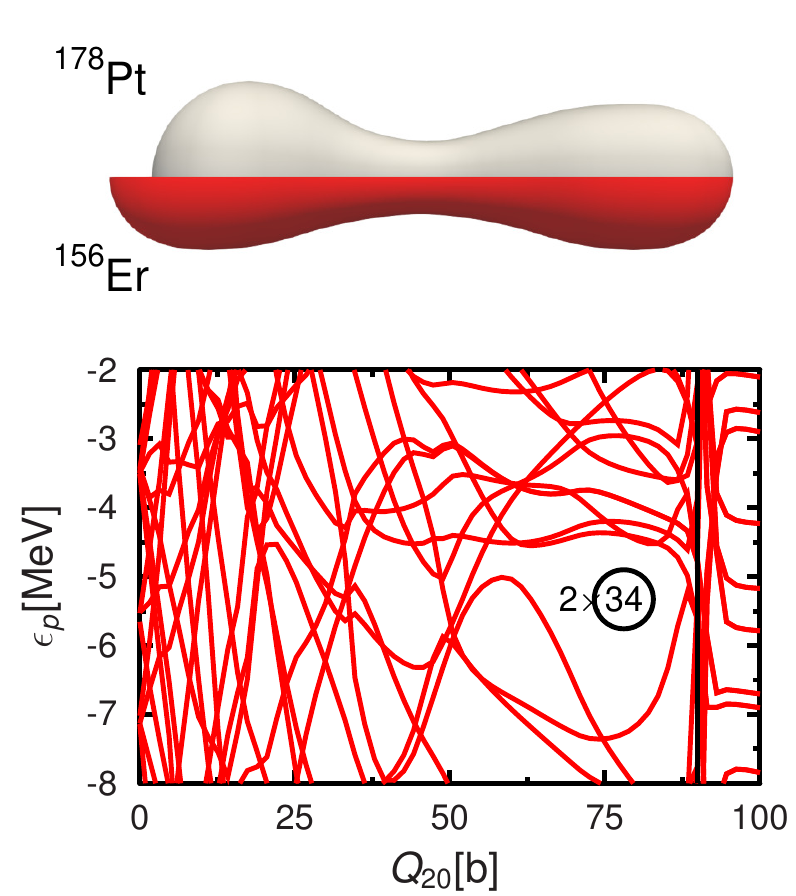}
\end{center}
\caption{Top: Comparison between the iso-density surface of the $^{178}$Pt at deformation $Q2$=80b and the $^{156}$Er at deformation  $Q2$=82b. Bottom: proton single-particle energies as a function of the quadrupole (lower left scale) deformation parameters in the $^{156}$Er fissioning system (Z~=~2~$\times$~34). The black vertical line marks the position of the scission configuration.  
} 
\label{fig:struc_Z36_N44_x2}
\end{figure}

\begin{figure*}[htb]
\begin{center}
\includegraphics[width=  0.9\linewidth]{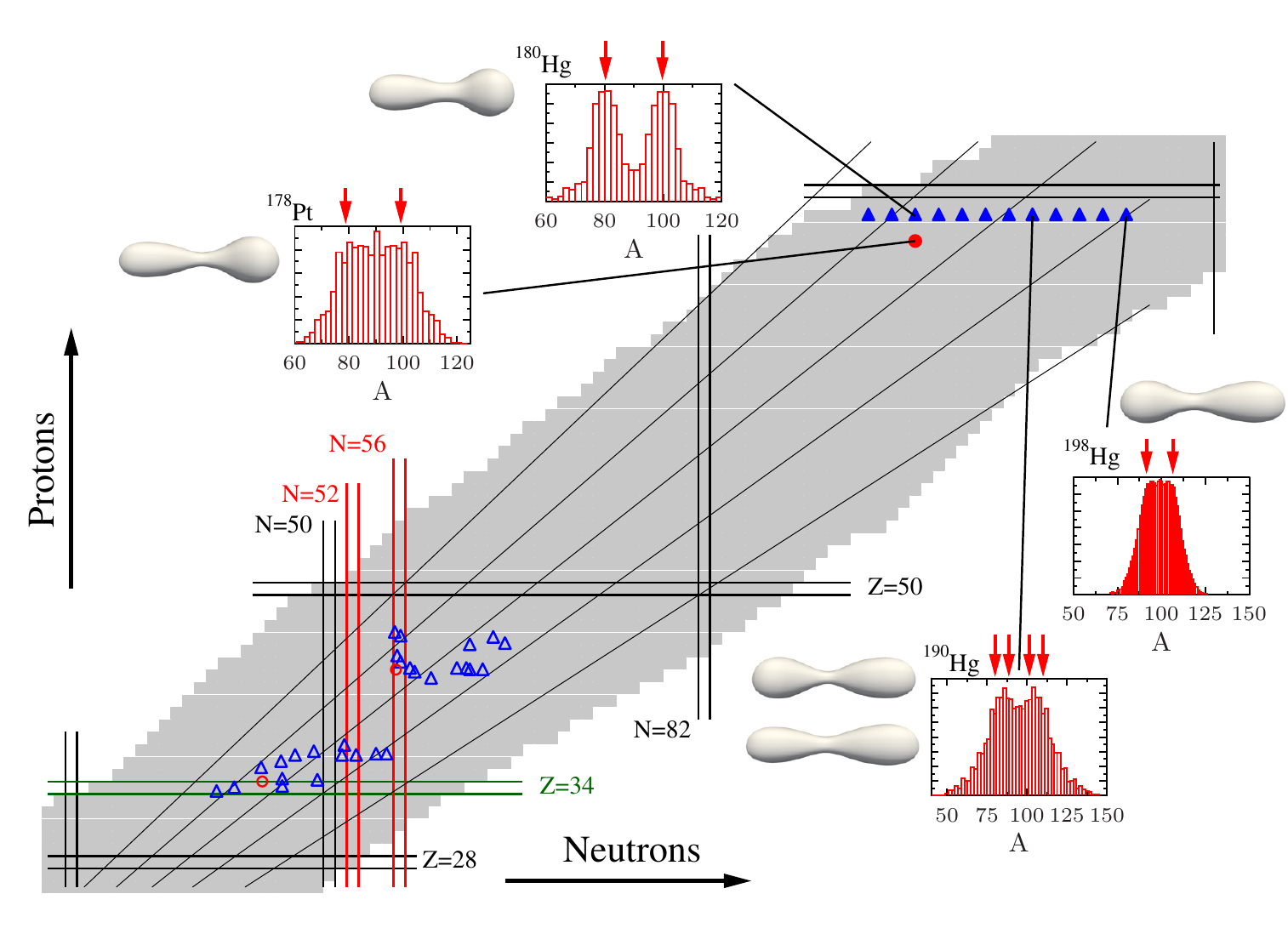}
\end{center}
\caption{ Nuclear chart with 
 spherical ``magic'' numbers (black solid lines) and main deformed shell gaps affecting fission of sub-lead nuclei (colored solid lines). 
Asymmetric fission properties of mercury isotopes (full triangles) and $^{178}$Pt (full circle) have been studied with constrained HF+BCS calculations. 
Expected centroids of light and heavy fragments are shown with associated open symbols. 
Insets show a comparison of theoretical predictions (arrows) with experimental data (histograms) from Refs.~\cite{itkis1989,andreyev2010,nishio2015,tsekhanovich2019}. Isodensities surface at half the saturation density $\rho_0/2=0.08$~fm$^{-3}$ at scission associated with each inset are shown for each asymmetric fission mode with the heavy fragment on the right.}
\label{fig:map}
\end{figure*}

We now introduce a complementary method to investigate shell effects in the pre-fragments. 
Instead of constraining the shape of an isolated fragment, the single-particle energies are studied in the entire fissioning nucleus to search for the appearance of energy gaps near scission. 
The main difficulty here is to assign the level to one fragment or to the other. 
However, this is not a problem if the fission is symmetric as in this case the shell gap is present in both fragments, e.g., if there is a gap  for $Z$ protons in the pre-fragment, we expect to see a gap for $2Z$ protons in the total system~\cite{simenel2014a}. 
Assuming that the shape of the pre-fragment does not depend strongly on the fissioning system, we can substitute an asymmetric fission $A\rightarrow B+C$ to a symmetric one $D\rightarrow B+B$.  This method has the adventage to disentangle the shell-structure of the two fragments B and C.

Figure~\ref{fig:struc_Z36_N44_x2} illustrates the method for the proton levels in the symmetric fission   $^{156}_{\,\,\,68}$Er$\rightarrow^{78}_{34}$Se$+^{78}_{34}$Se.
The presence of a shell gap at $2\times34$ indicates a possible enhancement of the formation of this fragment, including in non-asymmetric fissions. Although the light fragment in $^{180}$Hg asymmetric fission has $Z=36$, it is possible that it gains some stability thanks to the $Z=34$ deformed gap. Note that the shape of the light pre-fragment in  $^{180}$Hg or $^{178}$Pt asymmetric fission and in $^{156}$Er symmetric fission are very similar, as shown by the iso-density surfaces in the top of Fig.~\ref{fig:struc_Z36_N44_x2}. Thus, they are expected to have a similar shell structure.

Asymmetric fissions of mercury isotopes with an even number of nucleons ranging from $A=176$ to $A=198$ have been studied with these techniques. 
The predicted average fragments are shown on the nuclear chart in Fig.~\ref{fig:map} (open triangles) and given in supplemental material.
Each of the most neutron deficient isotopes $^{176-184}$Hg produces a compact heavy fragment with $N_H\simeq56$, while each of the less neutron deficient isotopes $^{190-198}$Hg
has a light compact fragment with $N_L\simeq51-55$ in which octupole shapes are favoured by the deformed gaps at $N=52$ and 56.
Note that these deformed shell effects are also expected to affect symmetric fission in $^{184-192}$Hg with production of fragments having $52$ to $56$ neutrons 
  (see, e.g., isodensity of $^{188}$Hg symmetric fission near scission in Supplemental Material Fig.~1).   
In addition to mercury isotopes, asymmetric fission of $A\simeq178$ isobars has also been studied and found to form compact heavy fragments 
with $N\simeq54-58$ neutrons, thus confirming the persistence of $N=56$ octupole deformed shell effects away from mercury isotopes (see supplemental material).

\begin{figure}[htb]
\begin{center}
\includegraphics[width= \linewidth]{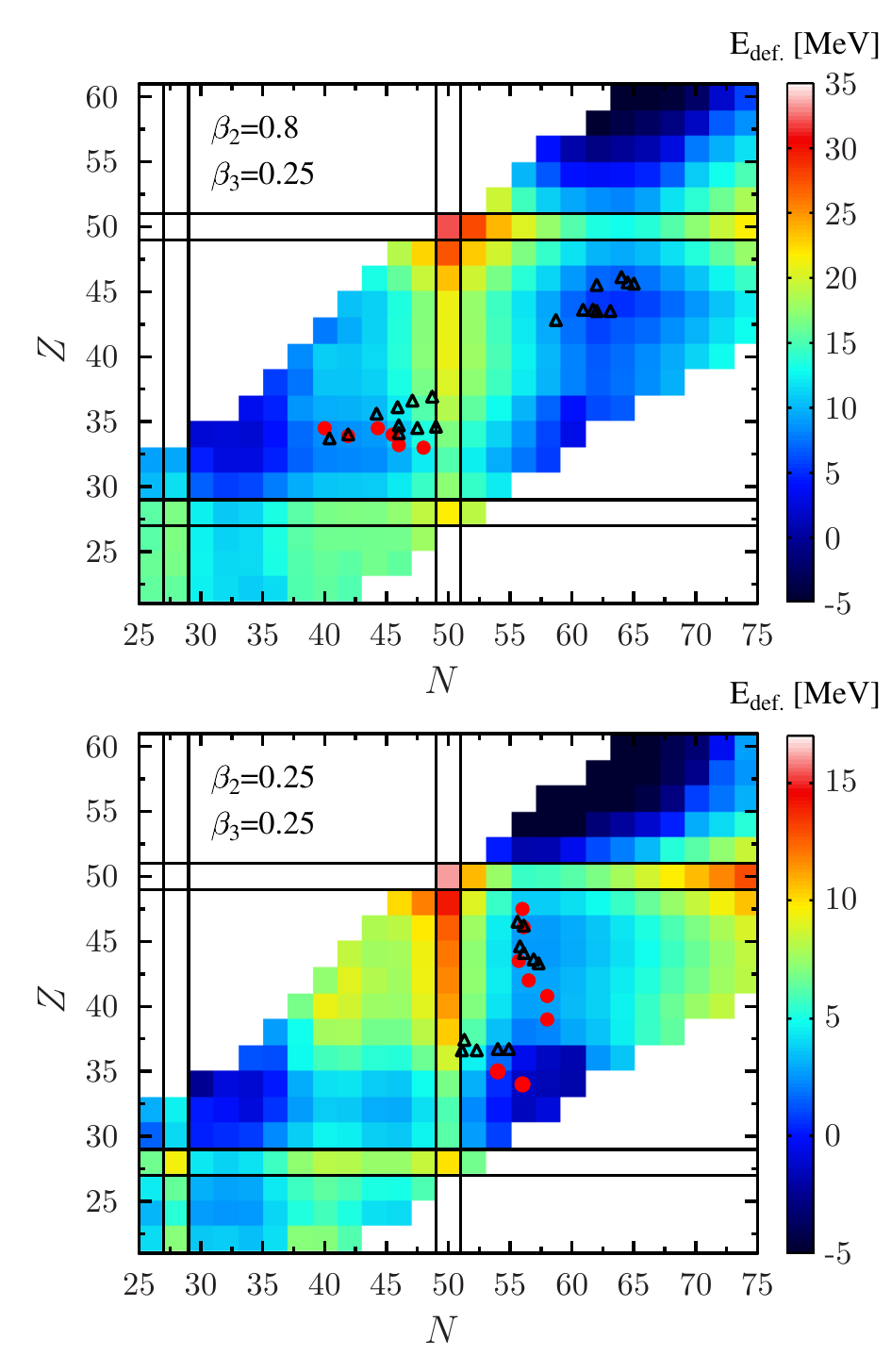}
\end{center}
\caption{ Deformation energy in even-even nuclei 
(background color)  for a constrained deformation corresponding to the elongated (top) and to the compact (bottom) shapes with  $\{\beta_{2}=0.8,\beta_{3}=0.25\}$ and $\{\beta_{2}=0.25,\beta_3=0.25\}$, respectively. The expected values of the neutron and proton numbers of the fission fragments in the asymmetric mode of systems from Supplemental Material table I (black triangles) and Supplemental Material table II (red dots) are shown if the fission fragments have approximatively the deformations used to generate the background figure. }
\label{fig:fig_def_x2}
\end{figure}

Elongated shell gaps are also observed that could influence the fission of mercury isotopes.  
The shell gap $Z=34$ (see Fig.~\ref{fig:struc_Z36_N44_x2}) is expected to play a role in the formation of the light fragments in $^{176-194}$Hg asymmetric fission. 
Interestingly, we found a second asymmetric valley in  $^{190,192,194}$Hg with  $Z_L\simeq34$ elongated fragments. All heavy fragments in $^{188-198}$Hg asymmetric fissions are elongated  with $Z_H\simeq42-46$ protons. This indicates a possible influence of quadrupole shell gap at $\beta_2>0.5$ with $42-46$ protons \cite{Macchiavelli1988,nazarewicz1985} 
(see Supplemental Material Fig.~3). 

On Fig.~\ref{fig:fig_def_x2}, we show the energy that is required to deform a nucleus from the spherical shape to the typical deformation of the fragments at scission in the case of the compact ($\beta_{2,3}=0.25$) and elongated ($\beta_{2}=0.8$ and $\beta_{3}=0.25$) shapes. The  positions of the fragments on the nuclear chart are found in the area for which the deformation energy is small.

Finally, our theoretical predictions are compared with available experimental mass distributions in $^{180,190,198}$Hg and $^{178}$Pt, shown in insets in Fig.~\ref{fig:map}.
All these nuclei exhibit various level of asymmetries in their experimental fission fragment mass distributions. 
Although more data are required to achieve definitive conclusions, it is encouraging to see that the predicted asymmetric modes are all compatible with the experimental distributions.


Asymmetric fission has been studied theoretically in the sub-lead region.
Its origin is interpreted as an effect of octupole correlations induced by deformed shell gaps at $N=52-56$ neutrons in the fission fragments (or pre-fragments), although shell gaps associated with large quadrupole deformations at 34 and 42-46 protons seem to also contribute.
Similar effects being present in actinide fission, we therefore conclude that the mechanisms driving mass-asymmetric fission are the same in both regions. 
It would be interesting to investigate the impact of octupole correlations in fission of superheavy elements (SHE), in which
 shell effects in $^{208}$Pb could induce superasymmetric fission \cite{poenaru2011,zhang2018b,santhosh2018,warda2018,matheson2019}.
Similar effects have also been predicted \cite{wakhle2014,umar2016,sekizawa2016,guo2018b,sekizawa2019} and observed \cite{wakhle2014,morjean2017} in quasifission. 
Despite being doubly-magic, $^{208}$Pb has a low-lying collective octupole vibrational state and could thus be formed as a fission fragment with a pear shape (as shown in supplemental material).


\begin{acknowledgments}
\textit{Acknowledgments} Discussions with D. J. Hinde, C. Schmitt, B. Jurado, A. Chatillon and  W. Nazarewicz are acknowledged. 
This work has been supported by the Australian Research Council under Grants No. DP160101254 and DP190100256. 
The calculations have been performed in part at the NCI National Facility in Canberra, Australia, 
which is supported by the Australian Commonwealth Government, 
 in part using the COMA system at the CCS in University of Tsukuba supported by the HPCI Systems Research Projects (Project ID hp180041), and using the Oakforest-PACS at the JCAHPC in Tokyo supported in part by Multidisciplinary Cooperative Research Program in CCS, University of Tsukuba. This work was supported by the Fonds de la Recherche Scientifique (F.R.S.-FNRS) and the Fonds Wetenschappelijk Onderzoek - Vlaanderen (FWO) under the EOS Project nr O022818F\\
\end{acknowledgments}

\section{Supplement material}

\subsection{Multipole moments and deformation parameters}

The quadrupole moment is expressed as 
 $$Q_{20}=\sqrt{\frac{5}{16\pi}}\int d^3r \,\rho(\mathbf{r}) (2z^2-x^2-y^2)$$
 and the octupole moment as 
 $$Q_{30}=\sqrt{\frac{7}{16\pi}}\int d^3r \,\rho(\mathbf{r}) [2z^3-3z(x^2+y^2)],$$
 where $\rho(\mathbf{r})$ is the density of nucleons. 
The $\beta_2$ and $\beta_3$ deformation parameters are obtained from the quadrupole and octupole moments following,
\begin{align}
\beta_{\lambda} = \frac{4 \pi }{3 A (r_0 A^{1/3})^{\lambda} }  Q_{\lambda 0},
\end{align}
with $r_0$~=~1.2~fm.

\subsection{Fermion localisation function} 

The localisation function is computed as \cite{becke1990,reinhard2011}
\begin{align}
	{ C}_{q \sigma}( {\bf r})  = \left[  1+ \left( \frac{\tau_{q \sigma } \rho_{q \sigma }  - \frac14 | \nabla \rho_{q \sigma } |^2  - {\bf j}^2_{q \sigma }  }{   \rho_{q \sigma }  \tau^{TF}_{q \sigma }  } \right)^2 \right]^{-1},
\end{align}
with the nucleon ($\rho_{q \sigma }$), kinetic ($\tau_{q \sigma }$) and current (${\bf j}_{q \sigma }$) densities defined as
\begin{align}
	\rho_{q \sigma } ({\bf r}) &= \sum_{\alpha \in q}  n_{\alpha}  \varphi^*_{\alpha} ({\bf r} \sigma)\varphi_{\alpha} ({\bf r} \sigma),\\
	\tau_{q \sigma } ({\bf r}) &= \sum_{\alpha \in q}  n_{\alpha} \left| \nabla \varphi_{\alpha} ({\bf r} \sigma) \right|^2, \\
	{\bf j}_{q \sigma } ({\bf r}) &= \sum_{\alpha \in q}  n_{\alpha}   {\rm Im} \left[  \varphi^*_{\alpha} ({\bf r} \sigma) \nabla \varphi_{\alpha} ({\bf r} \sigma) \right], 
\end{align}
where $q$ stands for neutron or proton and $\sigma$ is the spin.
$\tau^{TF}$ is the Thomas-Fermi approximation of the kinetic density.
To study the inner core of the nuclei, we suppress the localisation function on the surface of the fragments by applying the transformation \cite{zhang2016},
\begin{align}
	{ C}_{q \sigma}( {\bf r})  \rightarrow { C}_{q \sigma}( {\bf r})  \frac{\rho_{q \sigma } ({\bf r})}{ {\rm max} \left[  \rho_{q \sigma } ({\bf r}) \right]}.
\end{align}
The neutron ($q=n$) localisation function is obtained by averaging over the spin $\sigma$.

Localisation functions, together with isodensities, are shown in Supplemental Material Fig.~\ref{fig:Hg_scission} for mercury isotopes near scission. 
All calculations are in the asymmetric valley except the bottom one for $^{188}$Hg which forms two pear-shaped compact fragments with $N=54$ neutrons.

\begin{figure}[htb]
\begin{center}
\includegraphics[width=  \linewidth]{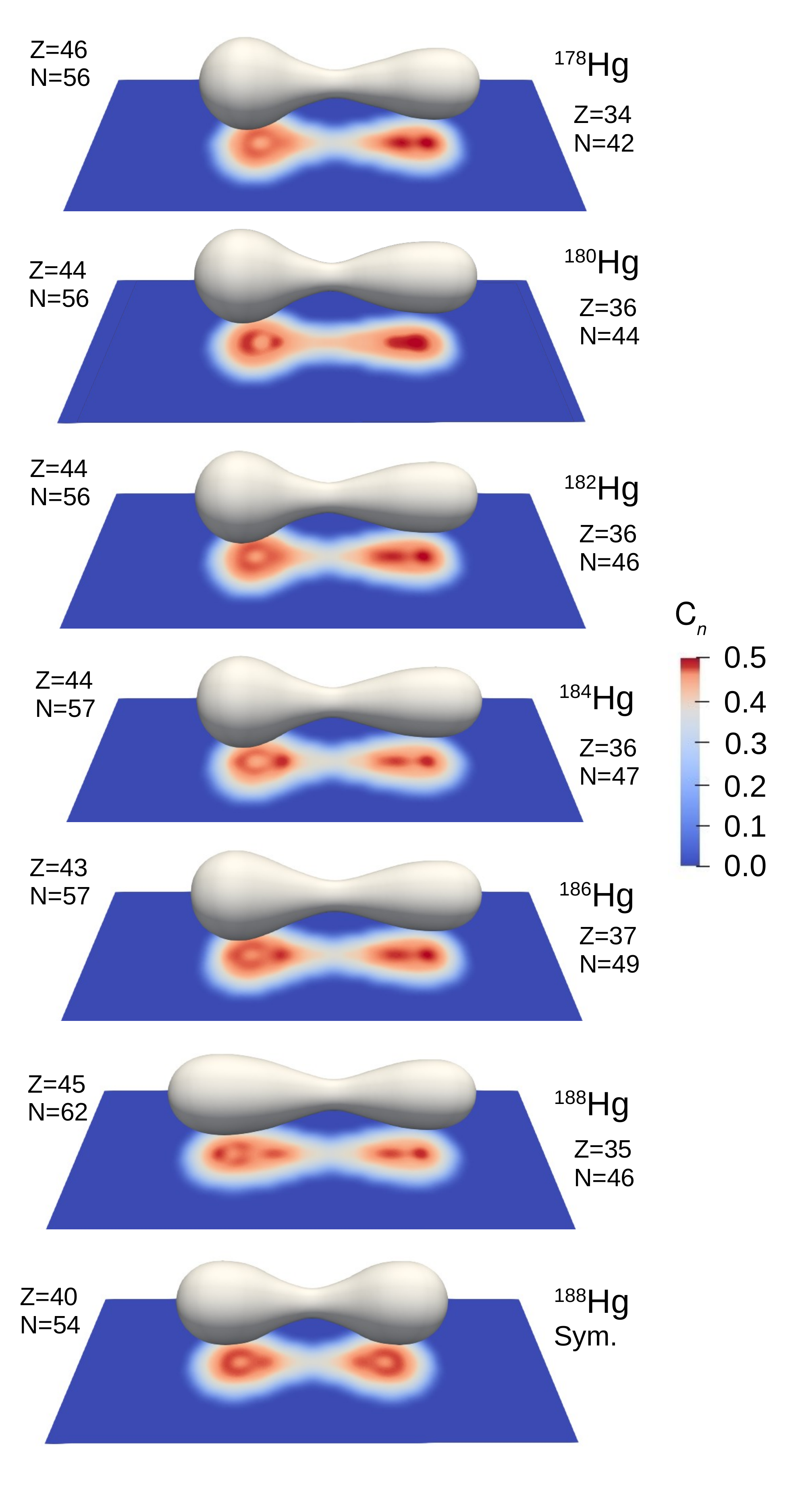}
\end{center}
\caption{ Isodensity surface at half the saturation density $\rho_0/2=0.08$~fm$^{-3}$  and neutron localisation function (projections) at scission in various mercury isotopes.  } 
\label{fig:Hg_scission}
\end{figure}

\subsection{Fragment properties}

The number of protons and neutrons in the fragments are determined from the proton and neutron densities on each side of the neck (defined by the plane between the fragments which has the smallest integrated density) just before scission. Scission is defined here as the configuration just before the system jumps from the fission valley to the fusion one when the $Q_2$ constraint is increased.
Although the results may in principle depend on the definition of the neck, we found that near scission, 
different definitions of the neck lead to less than one unit change of $Z$ and $N$ in the fragments. For instance, we checked that by considering a range of quadrupole deformations of varying by  20~b before scission, the variation of the number of protons and neutrons are found to be less than 1-2 units in the elongated modes, and much less than 1 unit in the case where the asymmetric path is well pronounced like in $^{180}$Hg.

The results for the fission of mercury isotopes and $A\simeq178$ isobars are given in tables \ref{tab:fragHg} and \ref{tab:frag178}, respectively.  
The fragments with a number of protons or neutrons associated with a deformed shell gap are indicated with colors (if their deformation matches the position of the shell gaps in Figs.~2 and~3 of the rapid communication). For clarity, only fragments with $N$ or $Z$ less than one unit away from the corresponding shell gap are colored. This criteria is of course  arbitrary. For instance, the compact heavy fragment in $^{186}$Hg asymmetric fission is likely to be influenced by $N=56$ despite the fact that it is found with 57.3 neutrons. 
Note also that the entire range $52-56$ is considered as a shell gap (e.g., the formation of a compact fragment with 54 neutrons is clearly favoured by both 52 and 56 gaps). 

\begin{table}[h]
 \caption{\label{tab:fragHg}
Number of protons and neutrons in the asymmetric fission fragments of $^{A}$Hg.
$N_H$, $Z_H$, $N_L$ and $Z_L$ are the numbers of neutrons and protons in the heavy and light fragments, respectively. The main deformation of the fragments, which is referred to as compact (Comp.) for $\beta_2\simeq0.2-0.4$ (boldface) or elongated (Elong.) for  larger $\beta_2$ (italic) is indicated for the heavy (def. H.) and light  (def. L.) fragments.
The numbers in red  are associated with compact deformed shell gaps $52-56$. 
The numbers in green and blue are associated with the elongated shell gap at $34$ and  $42-46$, respectively. See Figs.~2 and~3 of the main document and Supplemental Figure 4 where these shell gaps are identified.}
 \begin{ruledtabular}
 \begin{tabular}{c|ccc|ccc}
$A$ & $N_H$ & $Z_H$ & def. H. & $N_L$ & $Z_L$ & def. L. \\
\hline
198 & 63.1 & {\it \color{blue} 43.4} & {\it \color{blue} Elong.} & {\bf \color{red} 54.9} & 36.6 & {\bf \color{red} Comp.}  \\
196 & 62 & {\it \color{blue} 43.4} & {\it \color{blue} Elong.} & {\bf \color{red} 54} & 36.6  &  {\bf \color{red} Comp.}   \\
194  & 61.7 & {\it \color{blue} 43.5} & {\it \color{blue} Elong.} & {\bf \color{red} 52.3} & 36.5  &  {\bf \color{red} Comp.}   \\
& 65 & {\it \color{blue} 45.5} & {\it \color{blue} Elong.} & 49 &  {\it \color{darkgreen}  34.5}  & {\it \color{darkgreen} Elong.} \\
192 & 60.9 & {\it \color{blue} 43.5} & {\it \color{blue} Elong.} & {\bf \color{red} 51.1} & 36.5 & {\bf \color{red} Comp.}  \\
 & 64.5 & {\it \color{blue} 45.6} & {\it \color{blue} Elong.} & 47.5 & {\it \color{darkgreen} 34.4} & {\it \color{darkgreen}Elong.} \\
190 & 58.7 & {\it \color{blue} 42.7} & {\it \color{blue} Elong.} & {\bf \color{red} 51.3} & 37.3  &  {\bf \color{red} Comp.}   \\
 & 64 & {\it \color{blue} 46} & {\it \color{blue} Elong.} &  {\it \color{blue} 46} & {\it \color{darkgreen} 34}  & {\it \color{darkgreen} Elong.} \\
188 & 62 & {\it \color{blue} 45.4} & {\it \color{blue} Elong.} &  {\it \color{blue} 46}  & {\it \color{darkgreen}34.6} &{\it \color{darkgreen} Elong.} \\
186 & { 57.3} & 43.2 &  { Comp. }  &  48.7 & { 36.8} & { Elong.} \\
184 & {\bf \color{red} 56.9} & 43.5 &  {\bf \color{red} Comp.}   & 47.1 & { 36.5}  & { Elong.} \\
182 & {\bf \color{red} 56.1} & 44 &  {\bf \color{red} Comp.}   & {\it \color{blue}45.9} & { 36}  & {\it \color{blue} Elong.} \\
180 & {\bf \color{red} 55.8} & 44.5 &  {\bf \color{red} Comp.}   & {\it \color{blue}44.2} & { 35.5}  & {\it \color{blue} Elong.} \\
178 & {\bf \color{red} 56.1} & 46.1 &  {\bf \color{red} Comp.}   & {\it \color{blue} 41.9} & {\it \color{darkgreen}33.9} & {\it \color{darkgreen}Elong.} \\
176 & {\bf \color{red} 55.6} & 46.4 &  {\bf \color{red} Comp.}   & 40.4 & {\it \color{darkgreen} 33.6}  & {\it \color{darkgreen}Elong.} \\
 \end{tabular}
 \end{ruledtabular}
 \end{table}

\begin{table}[h]
 \caption{\label{tab:frag178}
Same as Tab.~\ref{tab:fragHg} for $A\simeq178$ isobars.
$N$ and $Z$ are the numbers of neutrons and protons of the fissioning nucleus.}
 \begin{ruledtabular}
 \begin{tabular}{cccccccc}
$N$ & $Z$ & $N_H$ & $Z_H$ & def. H. & $N_L$ & $Z_L$ & def. L. \\
\hline
96 & 82 & {\bf \color{red} 56} & 47.5 & {\bf \color{red} Comp.}  & 40 & {\it \color{darkgreen}34.5} & {\it \color{darkgreen}Elong.} \\
98 & 80 &  {\bf \color{red} 56.1} & 46.1 & {\bf \color{red} Comp.}  & {\it \color{blue}41.9} & {\it \color{darkgreen}33.9} & {\it \color{darkgreen}Elong.} \\
100 & 78 & {\bf \color{red} 55.7} & 43.5 & {\bf \color{red} Comp.}   & {\it \color{blue}44.3} & {\it \color{darkgreen}34.5}  & {\it \color{darkgreen}Elong.} \\
102 & 76 & {\bf \color{red} 56.5} & 42 &{\bf \color{red} Comp.}   &  {\it \color{blue} 45.5} & {\it \color{darkgreen}34}  &{\it \color{darkgreen} Elong.} \\
104 & 74 & 58 & 40.8 & {Comp.}  & {\it \color{blue} 46} & {\it \color{darkgreen}33.2}  & {\it \color{darkgreen} Elong.} \\
106 & 72 & 58 & 39 & {Comp.}  & 48 & {\it \color{darkgreen}33}  & {\it \color{darkgreen}Elong.} \\
108 & 70 & {\bf \color{red} 54} & {\bf \color{red} 35} &{\bf \color{red} Comp.}   &  {\bf \color{red} 54 }& {\bf \color{red} 35}  &{\bf \color{red} Comp.}   \\
112 & 68 & {\bf \color{red} 56} & {\bf \color{red} 34} & {\bf \color{red} Comp.}  &  {\bf \color{red} 56} &  {\bf \color{red}  34}  &{\bf \color{red} Comp.}   \\
 \end{tabular}
 \end{ruledtabular}
 \end{table}

The results for $A\simeq178$ nuclei are provided for completeness. 
The calculations were performed to investigate the evolution of the influence of  the $N=56$ octupole shell gap in the formation of the fragments. 
We see in Tab.~\ref{tab:frag178} that most of these nuclei indeed produce asymmetric fission fragments with $N_H\simeq56$. 
It is interesting to note that the $A=178$ isobars with $Z=72$ and 74 protons form heavy fragments with slightly more neutrons ($N_H\simeq58)$, which could be interpreted as an effect of a repulsion at $N=50$ in the light fragment due to the difficulty for these magic fragments to acquire the octupole deformation required at scission. 
Although only the most neutron deficient of these $A=178$ isobars could potentially be studied experimentally without requiring large excitation energy (which may wash out shell effects), these theoretical calculations confirm the influence of $N=56$ deformed shell gaps in the formation of the fragments. 
 
\subsection{Comparison with time-dependent calculations}

In order to test the pertinence of the static approximation for the determination of the fragments mass and charge in Tables \ref{tab:fragHg} and \ref{tab:frag178} we performed dynamical calculations with the TDHF+BCS method starting at $Q_{20}$ values close to (but before) scission. For the $^{180}$Hg, starting with a quadrupole deformation of $Q_{20}$=90b in the asymmetric fission valley gives a heavy fragment with N=55.7 and Z=44.6, in excellent agreement with the static calculation (see table \ref{tab:fragHg}). 
Both calculations reproduce well the centroids of the $^{180}$Hg experimental fragment mass distributions. For the $^{184}$Hg, a deviation of  0.6 protons and 0.2 neutrons is found starting with $Q_{20}$=100b. For the $^{186}$Hg which is a system in transition between two modes, starting at $Q_{20}$=120b, proton and neutron numbers are found to be Z=44.3 and N=57.1 with a deviation of 1.1 protons from the static case. These results support the choice of a static approximation for the determination of the average mass and charge in the fragments in fission of sub-lead nuclei, with a maximum variation with respect to dynamical calculations of the order of one nucleon.

\subsection{Transition of modes around $^{186}$Hg }

As discussed in the rapid communication, a transition is observed between the characteristics of asymmetric fission of $^{176-184}$Hg 
and $^{190-198}$Hg in which $N=52-56$ shell gaps affect the heavy and light fragments, respectively. 
We indeed observe in  Supplemental Material Fig.~\ref{fig:Hg_scission} a transition of modes around $^{186}$Hg, 
from a compact heavy fragment in $^{182}$Hg to an elongated one in $^{188}$Hg.
This transition can be explained by  effects related to the shell gaps identified in the present contribution.
In particular, isotopes around the $^{180}$Hg can fission asymmetrically into a heavy fragment with $N=56$ and a light one with $Z=34-36$. 
However, this is not the case for mercury elements heavier than $^{184}$Hg because of the (approximate) conservation of the $N/Z$ ratio.

\begin{figure}[htb]
\begin{center}
\includegraphics[width=0.8  \linewidth]{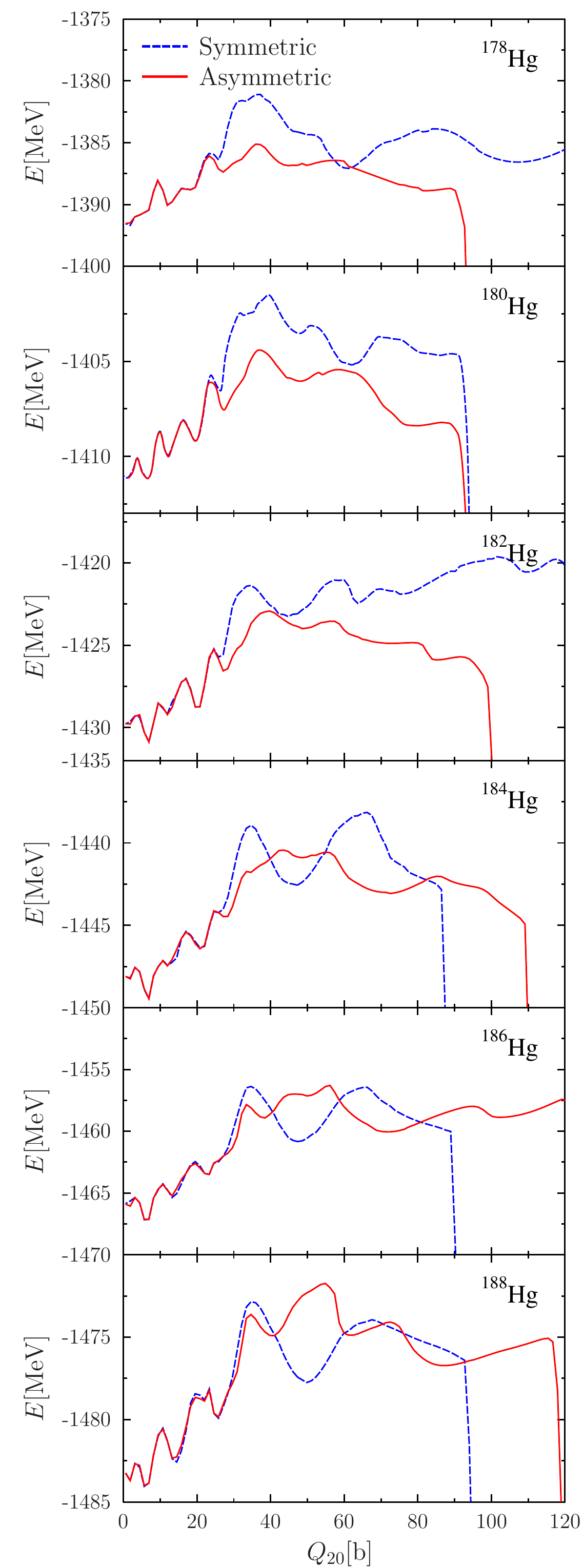}
\end{center}
\caption{ Potential energy as a function of quadrupole moment along the symmetric (dashed line) and asymmetric (solid line) fission paths of mercury isotopes.} 
\label{fig:PES_sym_asym_Hg}
\end{figure}

We also note in the rapid communication that this transition occurs in the region of mercury isotopes which can form symmetric fission fragments with $N=52-56$. 
The impact of these deformed shell effects in the fragments can be seen in the comparison between symmetric and asymmetric fission paths of $^{178-188}$Hg 
shown in Supplemental Material Fig.~\ref{fig:PES_sym_asym_Hg}.
The most neutron deficient mercury isotopes have lower asymmetric fission valleys, 
which can be interpreted as an effect of $N=52-56$ and $Z=34$ deformed shell effects.
On the contrary, the symmetric mode of isotopes around $^{180}$Hg fissions into two magic nuclei with $N\simeq50$ 
which is expected to increase the energy of this mode at  scission due to the difficulty for $N=50$ fragments to acquire an octupole shape.  
The $^{184-188}$Hg isotopes, however, have similar paths for both symmetric 
and asymmetric modes as $N=52-56$ deformed shell effects now lower the energy of the symmetric path. 
The impact of these shell effects can also be seen on the symmetric scission configuration of $^{188}$Hg 
(see bottom of Supplemental Material Fig.~\ref{fig:Hg_scission}) which exhibits compact octupole deformed fragments.

\subsection{ Shell structure of superdeformed $^{104}$Ru }

\begin{figure}[h]
\begin{center}
\includegraphics[width= 0.8 \linewidth]{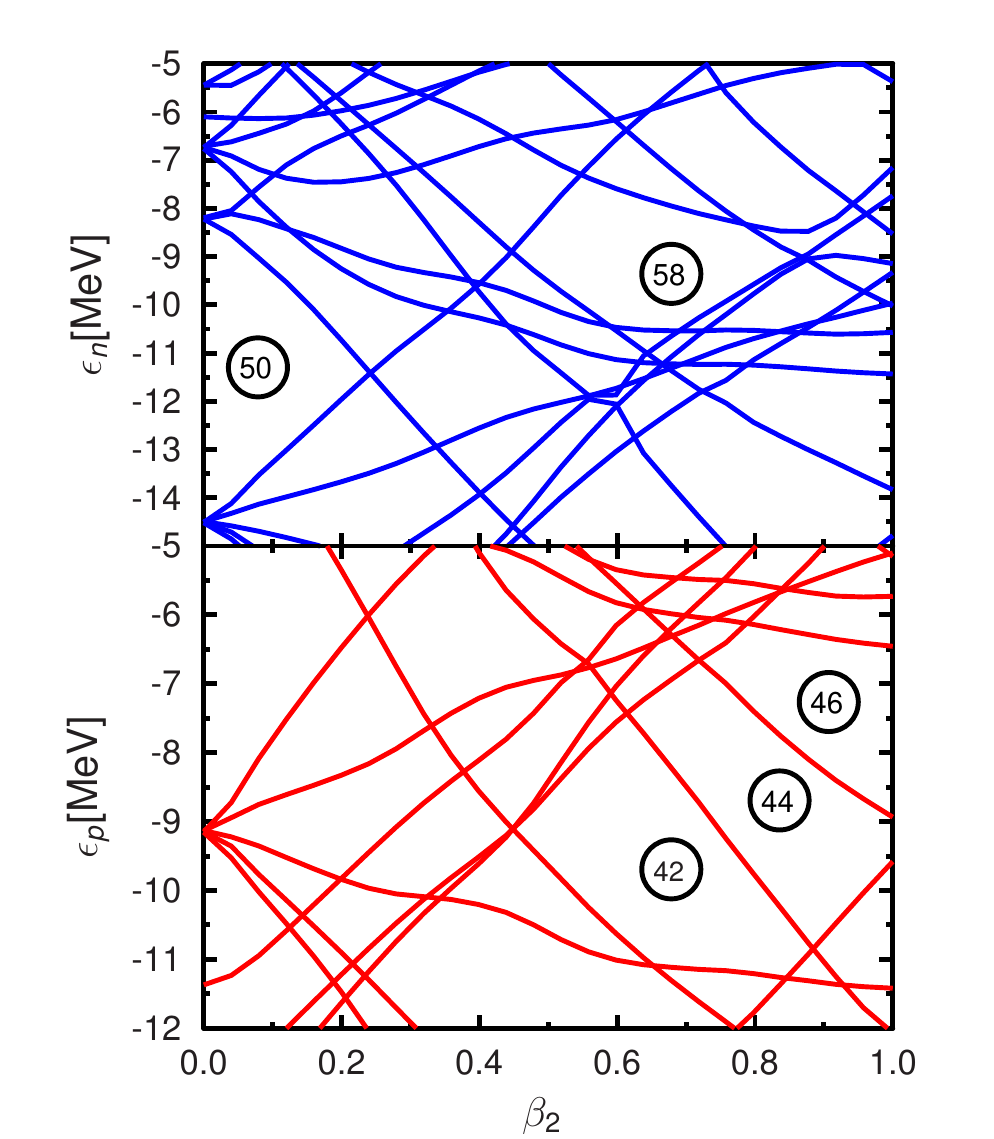}
\end{center}
\caption{ Neutron (top) and proton (bottom) single-particle energies as a function of the quadrupole  deformation parameters in $^{104}$Ru. }
\label{fig:struc_Ru104}
\end{figure}

The superdeformed shell gaps for nucleon number 42,44 and 46 are already well described in the literature
\cite{Macchiavelli1988,nazarewicz1985}. On Supplemental Material Fig.~\ref{fig:struc_Ru104}, we confirm the presence of those deformed gaps with the Sly4d functional.

\subsection{$^{294}$Og superasymmetric fission}

\begin{figure}[htb]
\begin{center}
\includegraphics[width=  \linewidth]{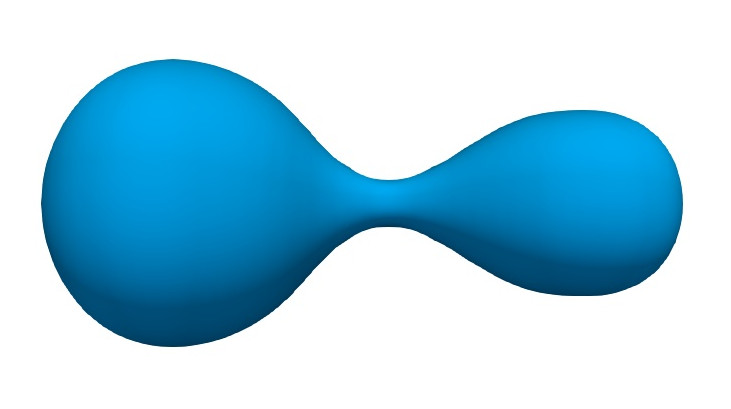}
\end{center}
\caption{Isodensity surface at half the saturation density $\rho_0/2=0.08$~fm$^{-3}$ of $^{294}$Og just before scission in the superasymmetric fission valley.} 
\label{fig:294Og}
\end{figure}

Preliminary time-dependent Hartree-Fock calculations with BCS dynamical correlations (see Ref. \cite{scamps2015a}  for details of the method) 
have been performed to study $^{294}$Og superasymmetric fission. 
An isodensity just before scission is shown in Fig.~\ref{fig:294Og}.
The heavy fragment (left) corresponds to a $^{208}$Pb doubly magic nucleus with an octupole deformation favoured by its low-lying $3^-$ state at an excitation energy of 2.6~MeV.

\bibliography{VU_bibtex_master}

\end{document}